\documentclass{article}

\usepackage[british]{babel}
\usepackage[useregional]{datetime2}
\DTMlangsetup[en-GB]{showdayofmonth=false}
\usepackage[numbers, sort]{natbib}
\usepackage[letterpaper,top=2cm,bottom=2cm,left=3cm,right=3cm,marginparwidth=1.75cm]{geometry}
\usepackage{amsmath}
\usepackage{amsfonts}
\usepackage{amssymb}
\usepackage{mathtools}
\usepackage{wrapfig}
\usepackage{graphicx}
\usepackage{tikz}
\usepackage[colorlinks=true, allcolors=blue]{hyperref}
\usepackage[page]{appendix} %

\usepackage{graphicx}
\usepackage{subcaption}
\usepackage{multicol}
\usepackage{lipsum}

\renewcommand{\vec}[1]{\boldsymbol{\mathbf{#1}}}
\renewcommand{\v}{\vec}

\newcommand{\sumin}{\sum_{i=1}^n}

\DeclareMathOperator*{\argmaxA}{arg\,max} %
\makeatletter
\newif\ifnobrackets
\renewcommand\@cite[2]{\ifnobrackets\else[\fi{#1\if@tempswa , #2\fi}\ifnobrackets\else]\fi\nobracketsfalse}
\newcommand\nbcite{\nobracketstrue\cite}
\makeatother
\usepackage{authblk}

\title{\vspace*{-2cm}{\textbf{\Large{Optimal Rebalancing in Dynamic AMMs}}}}
\author{Matthew Willetts}
\author{Christian Harrington}
\affil{QuantAMM.fi\vspace{-1mm}}
\begin{document}
\maketitle

\begin{abstract}
Dynamic AMM pools, as found in Temporal Function Market Making, rebalance their holdings to a new desired ratio (e.g. moving from being 50-50 between two assets to being 90-10 in favour of one of them)
by introducing an arbitrage opportunity that disappears when their holdings are in line with their target.
Structuring this arbitrage opportunity reduces to the problem of choosing the sequence of portfolio weights the pool exposes to the market via its trading function.
Linear interpolation from start weights to end weights has been used to reduce the cost paid by pools to arbitrageurs to rebalance.
Here we obtain the \emph{optimal} interpolation in the limit of small weight changes (which has the downside of requiring a call to a transcendental function) and then obtain a cheap-to-compute approximation to that optimal approach that gives almost the same performance improvement.
We then demonstrate this method on a range of market backtests, including simulating pool performance when trading fees are present, finding that the new approximately-optimal method of changing weights gives robust increases in pool performance.
For a BTC-ETH-DAI pool from July 2022 to June 2023, the increases of pool P\&L from approximately-optimal weight changes is $\sim25\%$ for a range of different strategies and trading fees.
\end{abstract}
\section{Introduction}

For AMM with fixed, unchanging trading functions, any change in the ratios between market prices of tokens in a pool lead to that pool holding less value than when the pool was capitalised.
So the market prices changing in ratio means the liquidity providers (in the absence of fees) are guaranteed to be outperformed by those who simply held their initial capital.

This problem of impermanent loss (IL), which can also be linked to Loss-versus-Rebalancing (LVR)~\cite{il_lvr,lvr}, is a serious challenge to AMM mechanisms where the trading function is static (which are the overwhelming majority of AMMs today).

In \emph{Temporal Function Market Making}~\cite{tfmm_litepaper}, AMM mechanisms are not used for core liquidity providing, but instead as an ultra-efficient rebalancing engine for running a time-varying portfolio.
This is done by having geometric mean market maker (G3M) pools where the weights are always changing from block to block following a chosen quantitative asset management strategy.

Arbitrage opportunities arising from changes in a pool's weights are how the pool pays for rebalancing to be done.
This can be interpreted as an auction~\nbcite[A.4]{tfmm_litepaper}, and if the pool's strategy for changing weights over time can provide greater P\&L than has to be paid to arbitrageurs to rebalance, then the pool's LPs can experience an uplift in the value of their holdings.

TFMM pools have a G3M trading function, for $N$ tokens with reserves $\v R=\{R_1, R_2,...,R_N\}$ we have:
\begin{equation}
    \prod_{i=1}^N R_i^{w_i(t)}= k(t), \quad\mathrm{where\,} \sum_{i=1}^N w_i(t) = 1, \,\,\mathrm{and\,\,} \forall i\,\, 0< w_i(t)<1,
    \label{eq:TFMM}
\end{equation}
with $\v w(t)=\{w_1(t), w_2(t),...,w_N(t)\}$ varying from block-to-block.
Trades still have to preserve (or increase) $k(t)$ at the time of that trade. (For simplicity we study the zero-fees case.)

In this paper, we explore the setting where we assume \emph{constant} market prices during the rebalancing process and \emph{changing} weights.
The question emerges: if we have initial weights $\v w(t_0)$ and the pool's strategy is such that we need to get to $\v w(t_f)$ (where $t_f$ is multiple blocks in the future) what is the optimal sequence of weights to incur the least amount of slippage for the pool, maximally reducing the pool's rebalancing cost?

We will first give a quick overview of how pools' reserves change when weights change, then describe how we can optimise this process, and then give simulated results demonstrating improved performance.

\newpage

\section{Weight Interpolation: Execution Management on TFMMs}
\label{sec:weight_interpolation}
Weight changes when the pool is in equilibrium---quoting the market price---creates an arbitrage opportunity.
This is desirable, we need to incentivise the pool's reserves to be re-balanced to be in line with the new weights.

We begin with some weights $\v w(t_0)$ and the pool in equilibrium.
Then a block later, at time $t' = t_0+\delta_t$, we have new weights $\v w(t')$ (but market prices $p$ do not change, so $\v p(t_0)=\v p(t')$).
This change in weights leads to a change in quoted prices.
For tokens with increasing weights, the pool is offering to buy the tokens from arbitrageurs for slightly more than their market price; for tokens with decreasing weights, the pool is offering to sell the tokens to arbitrageurs for slightly less than their market price.
In this way, the pool pays arbitrageurs to rebalance to the pool's desired holdings.

After this rebalancing arbitrage trade has happened, the pool holds reserves
\begin{equation}
    \v R(t') = \v R(t_0) \frac{\v w(t')}{\v w(t_0)}\prod_{i=1}^N \left(\frac{w_i(t_0)}{w_i(t')}\right)^{w_i(t')}.
\label{eq:reserve_change_weights}
\end{equation}
See Appendix~\ref{app:tfmm_update} for the derivation.

Of course, it is desirable for LPs for the arbitrage opportunity to be smaller rather than larger.
This is roughly analogous to `execution management' in TradFi, where a trader wishes for their transactions to follow best execution policies that reduce cost.
A simple way to reduce the arbitrage opportunity from a weight update is to spread out the weight update over a period in time.

In other words, when weight updates are divided up into a series of smaller steps the total resulting arbitrage from all of those multiple steps is less than if the update were done in one step, under the assumption that market prices are constant over the period of time in which weights are changing.

\subsection{A family of beneficial interpolations}
\label{ssec:broad_family_of_possible_interpolations}
We can compare $\v R^{1{\text -}\mathrm{step}}$, the reserves in the pool when we have directly updated the weights from $\v w(t_0)\rightarrow\v w(t_0) +\Delta \v w$, a one-step process, to $\v R^{2{\text -}\mathrm{step}}$, the reserves when we have done this weight update via a two-step process: $\v w(t_0) \rightarrow \Tilde{\v w} \rightarrow \v w(t_0) + \Delta \v w$, where $\forall i,\,\Tilde{w}_i \in [w_i(t_0),w_i(t_0)+\Delta w_i]$ (as well as $0<\Tilde{w}_i<1$ and $\sum_{i=1}^N \Tilde{w}_i=1$).
We show in Appendix \ref{app:general_interpolation} that any member of this family of intermediate values for $\Tilde{\v w}$ gives a lower arbitrage cost.

\paragraph{Example: Linear Interpolation}
We can show this result for the particular case of linear interpolation via Eq~\eqref{eq:reserve_change_weights}.
Here the two step process is: $\v w(t_0) \rightarrow \v w(t_0) + \frac{1}{2}\Delta \v w \rightarrow \v w(t_0) + \Delta \v w$.
This technique is used for liquidity bootstrap pools and stochastic changing CFMMs. 
We find that
\begin{align}
    \v R^{2{\text -}\mathrm{step}}_{\mathrm{linear}} = \v R^{1{\text -}\mathrm{step}}_{\mathrm{linear}}\prod_{j=1}^N \left(1+\frac{\Delta w_j}{2 w_j(t_0)}\right)^{\frac{\Delta w_j}{2}}.
    \label{eq:arb-bisection}
\end{align}
We can get this result for linear interpolation via Eq~\eqref{eq:reserve_change_weights}.
See Appendix~\ref{app:tfmm_update} for the derivation. 
The elementwise ratio between $\v R^{2{\text -}\mathrm{step}}_{\mathrm{linear}}$ and $\v R^{1{\text -}\mathrm{step}}_{\mathrm{linear}}$ is always greater than or equal to one.
This can be seen trivially.
When $\Delta w_j$ is $>0$, its term in the product is $>1$, as we have a number $>1$ raised by a positive power.
When $\Delta w_j$ is $<0$ its term in the product is again $>1$, as we have a number $<1$ raised by a negative power.
(And when $\Delta w_j$ is $=0$ the term in the product is $1$, as we have $1^0=1$.)

This means that for any weight change to be done in one leap, we can produce a cheaper weight update procedure by going through and equilibrating at the midway value.
After we apply this `bisection' once, we can apply the same argument again on each half of the new trajectory, dividing changes in weights in half again.

Informally, this tells us that we want weight changes to be maximally `smooth'.
For given start and end weights, choosing to linearly interpolate the value of the weights in the pool at block-level resolution is a simple and effective (and gas-efficient) way to give us good weight changes.

\newpage

\section{Non-linear weight interpolation schemes}
While it is always better to perform a weight change via linear interpolation than to do it in one go, the benefit `caps out' at the block resolution.
Once one has hit that limit, is there anything else one can do?
We will show that there are even more capital-efficient ways for pools to offer arbitrage opportunities to the market.

For a sequence of $f+1$ weights, indexed from $t_0$ to $t_f$, $\{\v w(t_k)\}_{k=0,...,f}$, and constant prices the final reserves are
\[\v R(t_f)=\v R(t_0) \frac{\v w(t_f)}{\v w(t_0)}\prod_{k=1}^f \prod_{j=1}^N \left(\frac{w_j(t_{k-1})}{w_j(t_{k})}\right)^{w_j(t_k)}, \]
which we obtain from repeated application of Eq~\eqref{eq:reserve_change_weights}.

Within this framework, one can attack this as an optimisation problem, as we are interested in finding the trajectory of weights that minimises the amount that is paid to arbs, i.e. maximises the pool value subject to constraints on the weights
\begin{align}
    \{\v w^*(t_k)\}_{k=1,...,f-1}={\argmaxA_{\{\v w(t_k)\}_{k=1,...,f-1}}}\Bigg[& \v p \cdot \v R(t_f) + \sum_{\ell=1}^{f-1} \left(\sum_{m=1}^N\mu_{\ell,m} w_m(t_\ell)\right)
    &- \sum_{k=1}^{f-1} \Lambda_k \left(\sum_{i=1}^N w_i(t_k) - 1\right)\Bigg],\label{eq:opt_w_interp}
\end{align}
which fulfil the KKT conditions with $\{\mu_{\ell,m}\}$ and $\{\Lambda_k\}$ as multipliers.
The optimum can be numerically calculated, as can more sophisticated approaches including changing prices (or even uncertain predictions of future prices) and the presence of fees.
These calculations could be run on-chain or otherwise made available to a TFMM pool.
During a TFMM weight update call is it possible to put in an external oracle call that returns a `weight trajectory array'.
Of course, there is a trade-off here in greater running costs vs the improvement from a more-complex weight trajectory.
The greater a pool's TVL and the cheaper it is to obtain an optimised weight trajectory (by on-chain calculation or from an oracle call), the easier it would be to justify this expenditure.

\subsection{Approximating the optimal weight interpolation}
\paragraph{Optimal intermediate weights for small weight changes}
Assuming constant market prices over the course of interpolation, here we will describe a cheap-to-compute approximation to the zero-fees solution of Eq~\eqref{eq:opt_w_interp} that provides a superior method of pool rebalancing over linear interpolation.

During the derivations for \S\ref{ssec:broad_family_of_possible_interpolations} in Appendix~\ref{app:general_interpolation} we defined $r$, the ratio of reserves between when doing a 2-step weight change ($\v w(t_0) \rightarrow \Tilde{\v{w}} \rightarrow \v w(t_f)$) and a direct 1-step weight change ($\v w(t_0) \rightarrow \v w(t_f)$).

\begin{equation}
r=\prod_{j=1}^N \frac{w_j(t_0)^{\Tilde{w}_j}}{w_j(t_0)^{w_j(t_f)}}\frac{\Tilde{w}_j^{w_j(t_f)}}{\Tilde{w}_j^{\Tilde{w}_j}}.
\label{eq:2stepbetter}
\end{equation}

If we assume that the weight change from $\v w(t_0) \rightarrow \v w(t_f)$ is small, we can directly optimise Eq~\eqref{eq:2stepbetter} without needing to impose the constraints of Eq~\eqref{eq:opt_w_interp}.
Doing this (Appendix \ref{app:optimal}), we find that the optimal intermediate value is
\begin{equation}
\Tilde{w}_i^* = \frac{w_i(t_f)}{W_0\left(\frac{e w_i(t_f)}{w_i(t_0)}\right)}
\label{eq:best_w_tilde}
\end{equation}
where $W_0(\cdot)$ is the principal branch of the Lambert W function and $e$ is Euler's constant.
The Lambert W function is a special function, the solution to a transcendental equation.\footnote{So not trivial to compute on-chain}
While there are methods to approximate $W_0(\cdot)$ (eg \cite{fastlambert,knuthlambert}) these a) can require (nested) calls to $\log(\cdot)$ that are not ideal for constrained-compute environments and b) are best over particular ranges of inputs while for us the ratios of start and end weights can take a broad range of values.
We chose a different approach that gives good performance.

\paragraph{Approximating the optimal value}
We can bound the value of $\Tilde{w}_i^*$ from above and below by the arithmetic mean and geometric mean (respectively) of $\{w_i(t_0), w_i(t_f)\}$:
\begin{equation}
 \sqrt{w_i(t_0)w_i(t_f)})<\Tilde{w}_i^*< \frac{1}{2} \left(w_i(t_0) + w_i(t_f)\right).
\label{eq:w_tilde_bound}
\end{equation}
See Appendix~\ref{app:bounding} for derivation.
We find that using the arithmetic mean \emph{of} the geometric\footnote{Note that $\lim_{|\v w(t_f)-\v w(t_0)|_1 \to 0} \left(\sumin\sqrt{w_i(t_0)w_i(t_f)}\right)=1$.} and arithmetic means of $\{w_i(t_0), w_i(t_f)\}$ provides a simple and effective method for approximating the optimal intermediate weight.\footnote{We could use another mean here (and in the final construction of $\breve{\v w}(t_k)$) geometric or harmonic say, but the arithmetic mean empirically works well and is cheaper to compute.}

\paragraph{From a two-step to multi-step approximation}
We can directly `bootstrap' the approximations derived above into giving close-to-optimal many-step weight trajectories, even though the base analytic result, Eq~\eqref{eq:best_w_tilde}, is only for a single intermediate point.
Along a trajectory of many intermediate weights, the optimal $\Tilde{w}_i$ at any stage in that trajectory will lie above the geometric mean of the prior and post weights and below their arithmetic mean.

The approximately-optimal $f-1$-step weight trajectory $\{\breve{\v w}(t_k)\}_{k=1,...,f-1}$ for going from $\v w(t_0)$ to $\v w(t_f)$ is made by taking the average of the linear interpolation between these values and the geometric interpolation between these values (see Appendix~\ref{app:bootstrap}):
\begin{align}
    w_i^{\mathrm{AM}}(t_k) & = (1-\frac{k}{f})w_i(t_0) + \frac{k}{f}w_i(t_f) \label{eq:approx_optimal_traj_am},\\
    w_i^{\mathrm{GM}}(t_k) & = {(w_i(t_0))^{(1-\frac{k}{f})}(w_i(t_f))^{\frac{k}{f}}},\label{eq:approx_optimal_traj_gm}\\
    \breve{w}_i(t_k) &= \frac{w_i^{\mathrm{AM}}(t_k)+w_i^{\mathrm{GM}}(t_k)}{\sum_{j=1}^N\left({w_j^{\mathrm{AM}}(t_k)+w_j^{\mathrm{GM}}}\right)} \label{eq:approx_optimal_traj},
\end{align}
where we explicitly are normalising our weight trajectories to enforce that they sum to 1 at each time step.

\begin{figure}[htbp]
\begin{minipage}{0.45\textwidth}
    \caption{Example weight interpolations done by a) linear interpolation, b) approximately-optimal (Eq~\eqref{eq:approx_optimal_traj}) and c) numerically-obtained optimal (Eq~\eqref{eq:opt_w_interp}), for a $N=3$ token pool with $\v w(t_0)=\{0.05, 0.55, 0.4\}$, $\v w(t_f)=\{0.4, 0.5, 0.1\}$ and $f=1000$. At this scale the differences between b) and c) are not easily apparent.}
\label{fig:traj}
\centering
    \begin{subfigure}{\textwidth}
         \centering
        \includegraphics[width=1\textwidth]{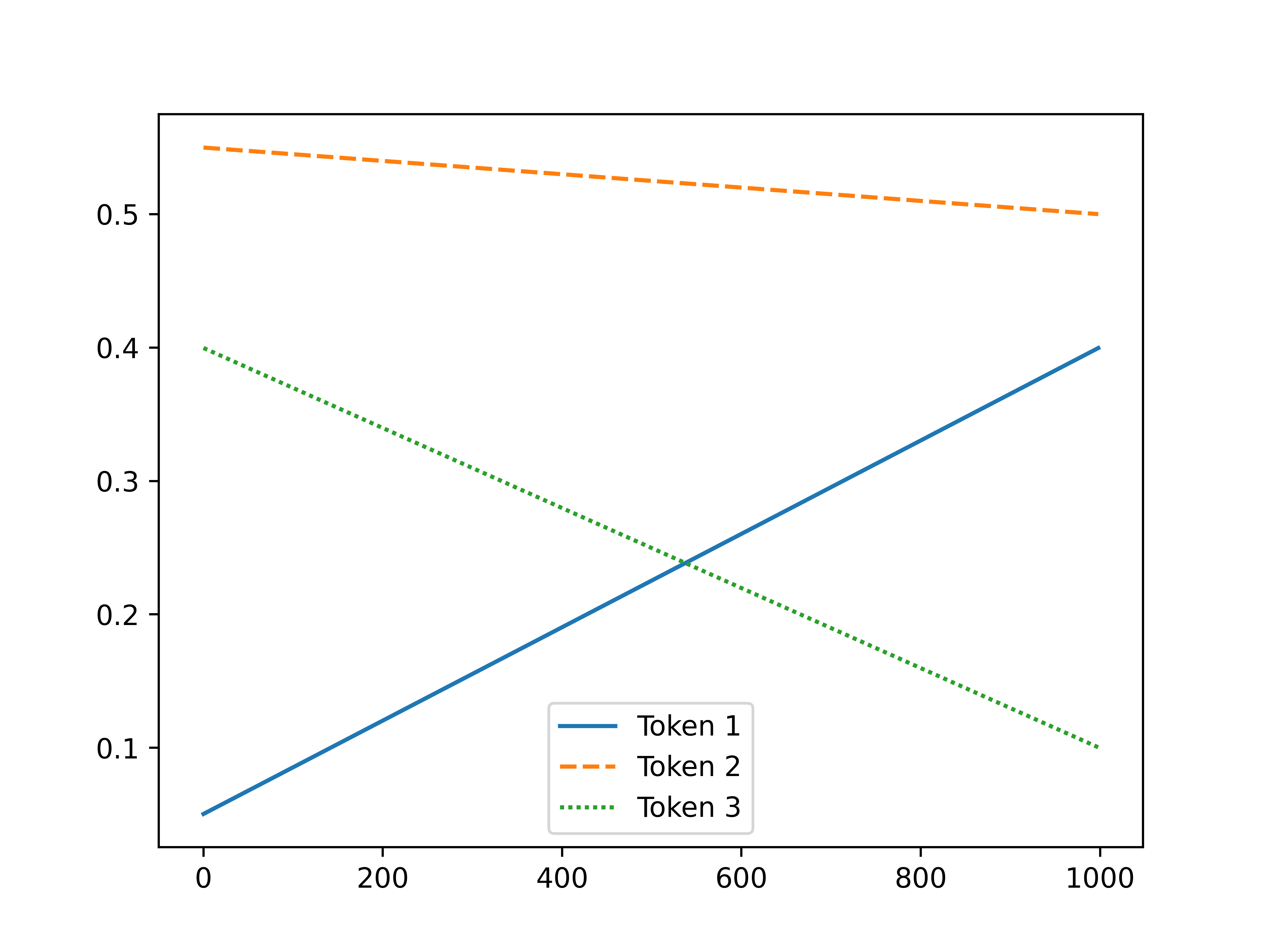}
        \caption{Linear}
        \label{fig:linear}
    \end{subfigure}
    \begin{subfigure}{\textwidth}
        \centering
        \includegraphics[width=1\textwidth]{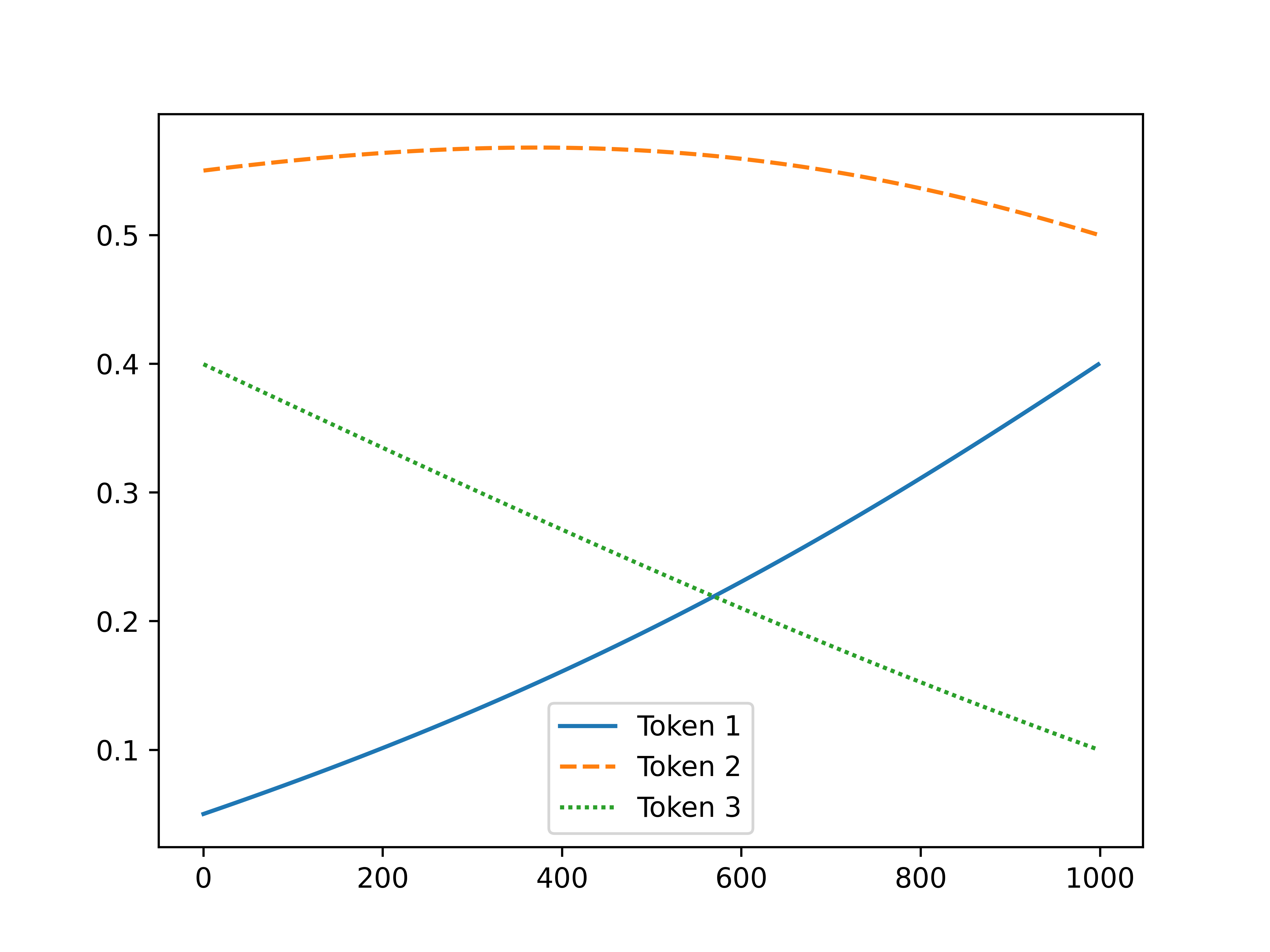}
        \caption{Approximately-optimal}
        \label{fig:approx}
    \end{subfigure}
    \begin{subfigure}{\textwidth}
        \centering
        \includegraphics[width=1\textwidth]{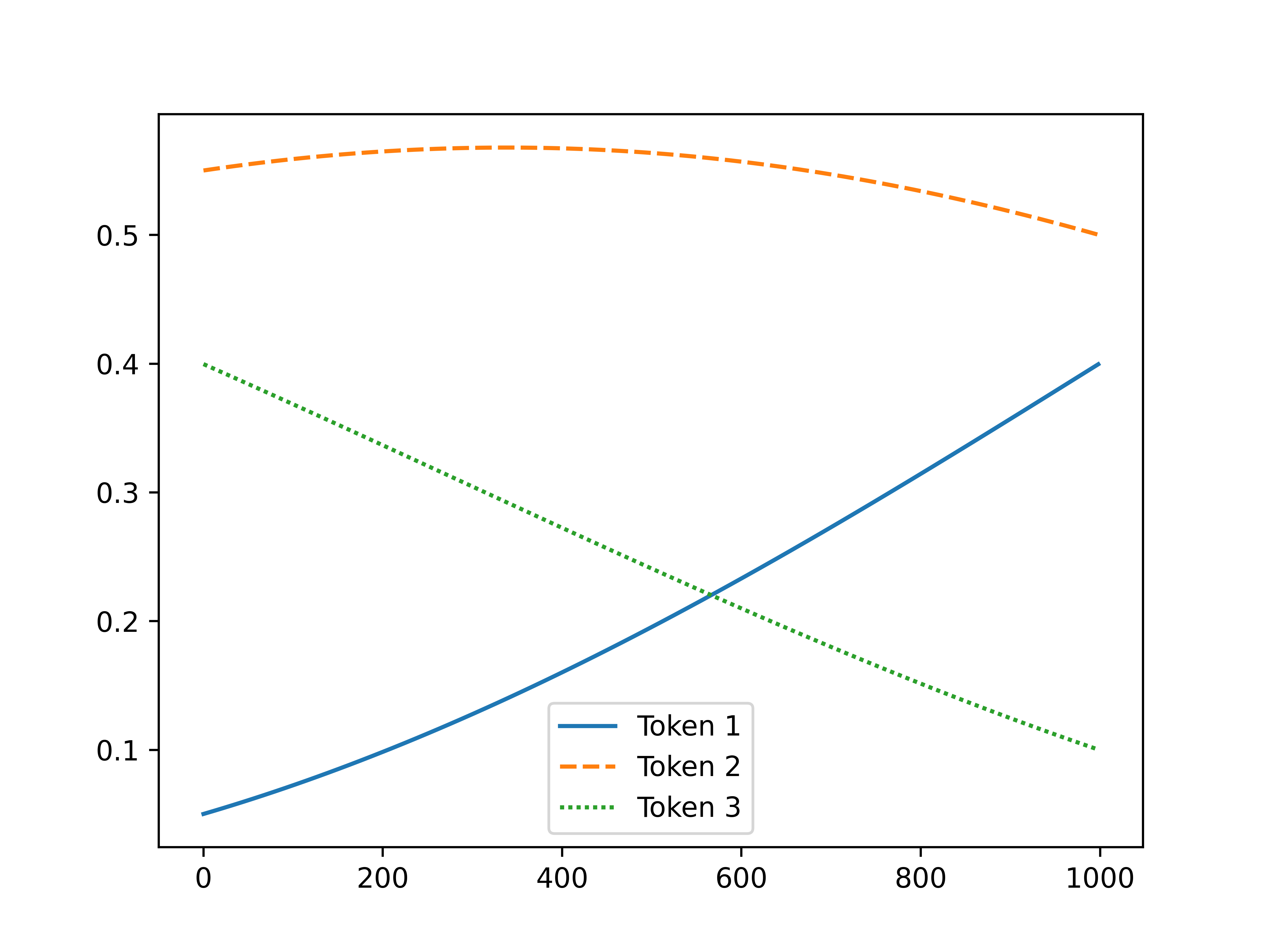}
        \caption{Optimal}
        \label{fig:optimal}
    \end{subfigure}
\end{minipage}\hfill
\begin{minipage}{0.45\textwidth}
    \vspace{0.1em}
    \caption{Plots showing the weight changes block to block (i.e. plotting each component of $\v w(t+1) - \v w(t)$) for a) linear interpolation, b) approximately-optimal (Eq~\eqref{eq:approx_optimal_traj}) and c) numerically-obtained optimal weights, same setup as Fig~\ref{fig:traj} above.}
    \centering
    \begin{subfigure}{\textwidth}
        \centering
        \includegraphics[width=1\textwidth]{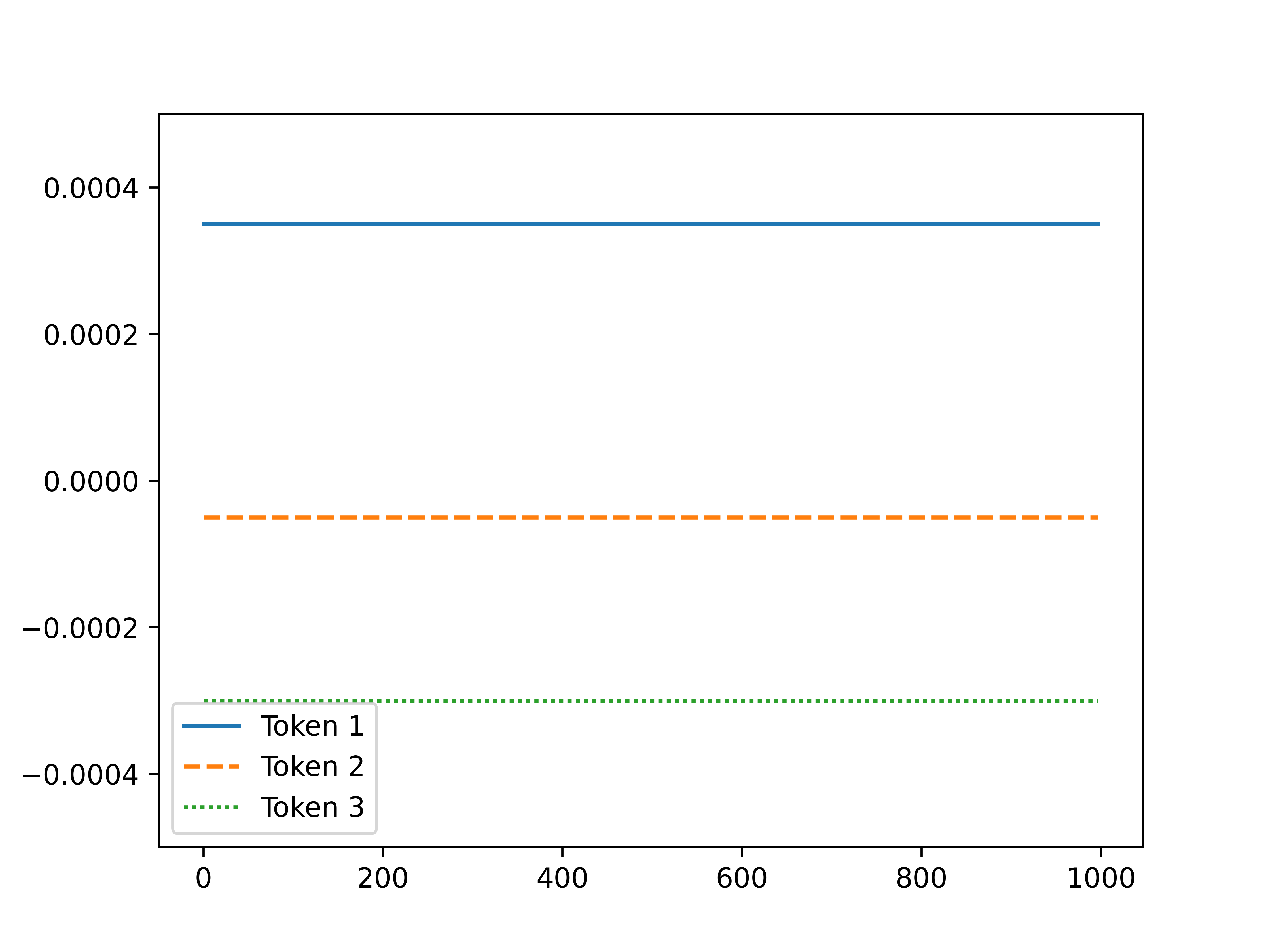}
        \caption{Linear}
        \label{fig:linear_change}
    \end{subfigure}
    \begin{subfigure}{\textwidth}
       \centering
        \includegraphics[width=1\textwidth]{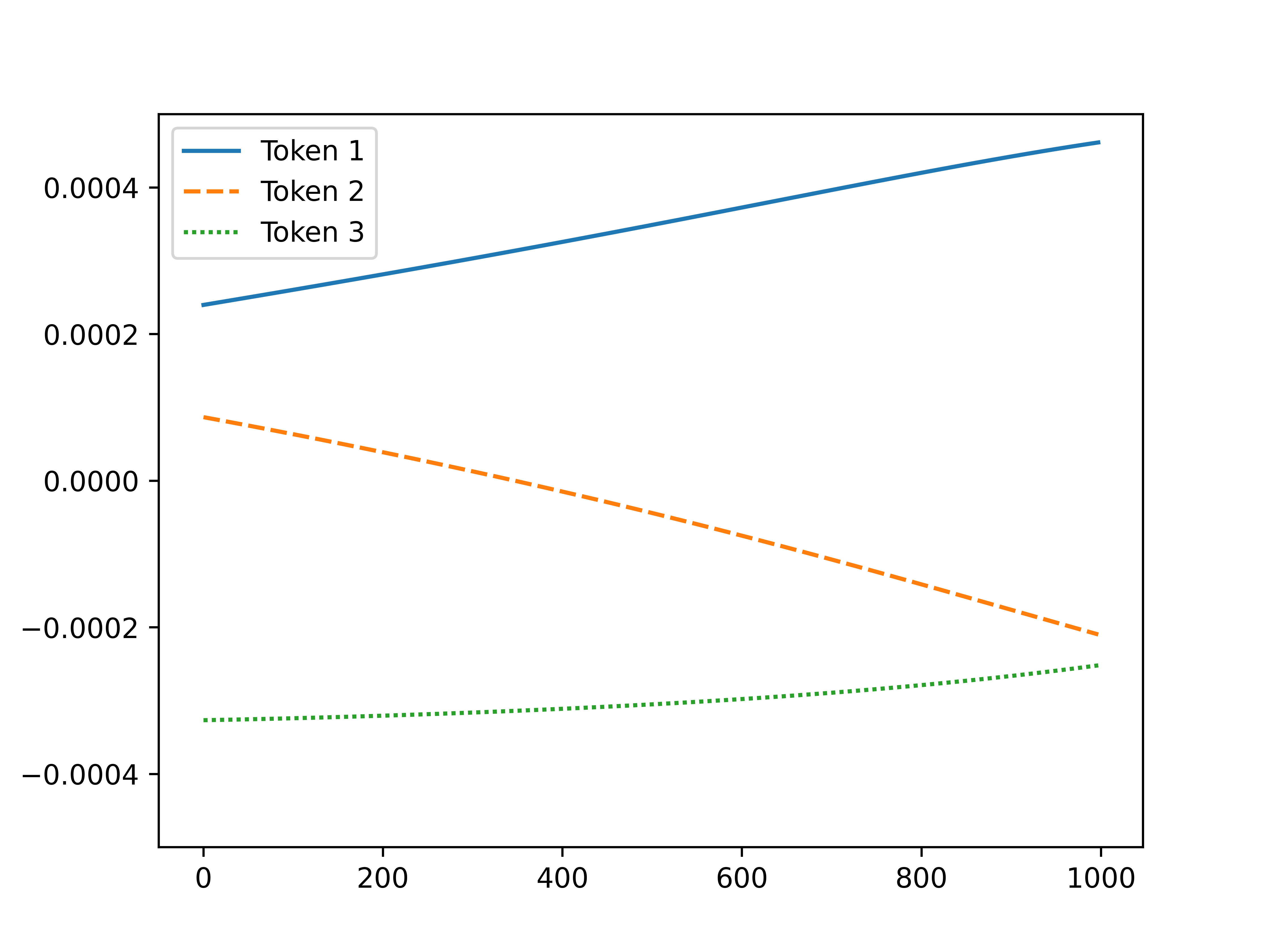}
        \caption{Approximately-optimal}
        \label{fig:approx_change}
    \end{subfigure}
    \begin{subfigure}{\textwidth}
        \centering
        \includegraphics[width=1\textwidth]{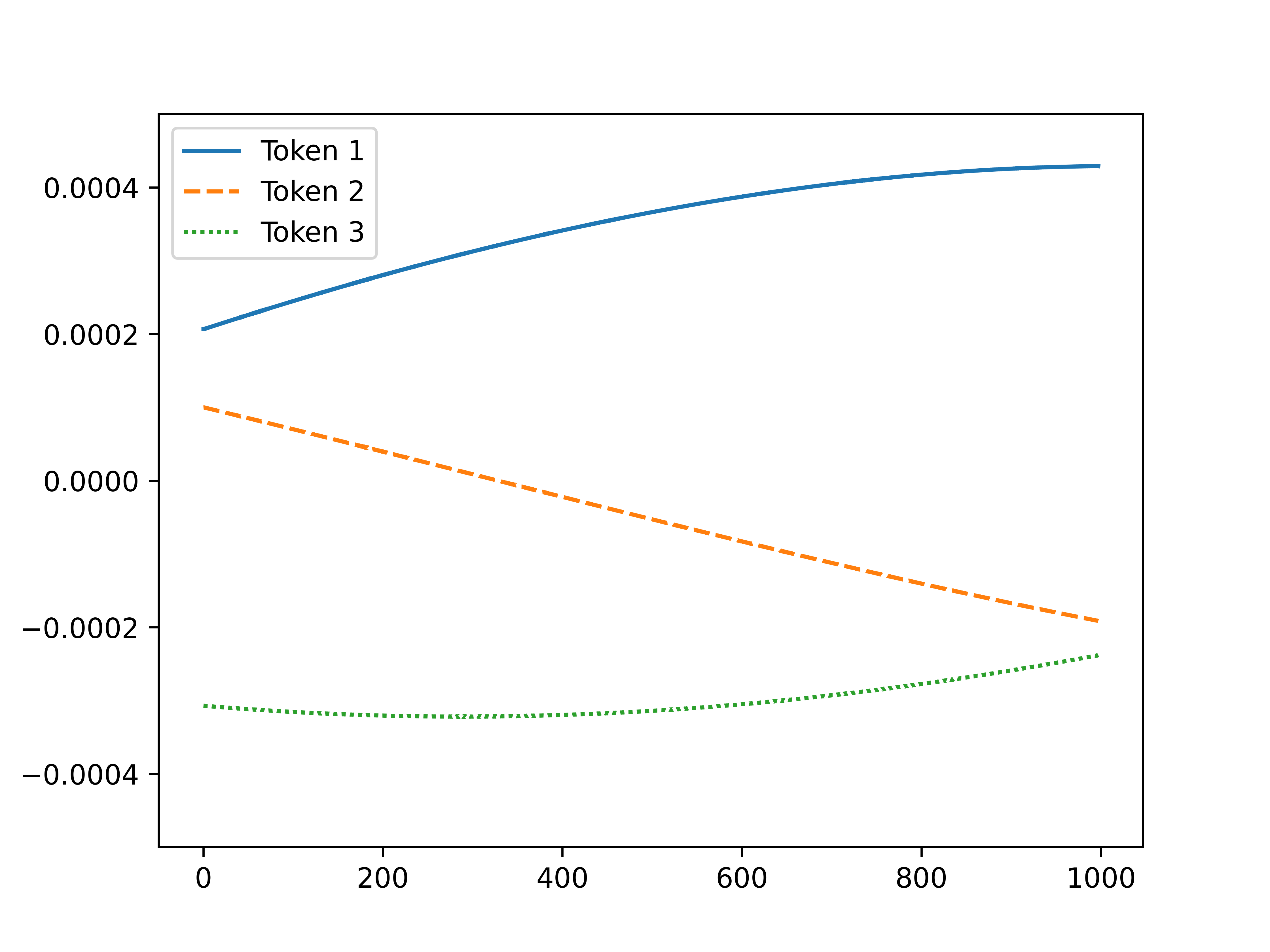}
        \caption{Optimal}
        \label{fig:optimal_change}
    \end{subfigure}
\label{fig:change}
\end{minipage}
\end{figure}

\begin{figure}[h]
    \caption{Plots showing how much linear interpolation (a) and approximately-optimal trajectories (b) deviate from the numerically-obtained trajectory $\{\v w^*(t_k)\}_{k=1,...,f-1}$, in the same setup as Fig~\ref{fig:traj}. We plot the differences. The approximately-optimal method very closely matches the numerically-obtained optimal.}
    \centering
    \begin{subfigure}{0.45\columnwidth}
    \centering
    \includegraphics[trim={0 0mm 0 0mm},clip,width=0.93\textwidth]{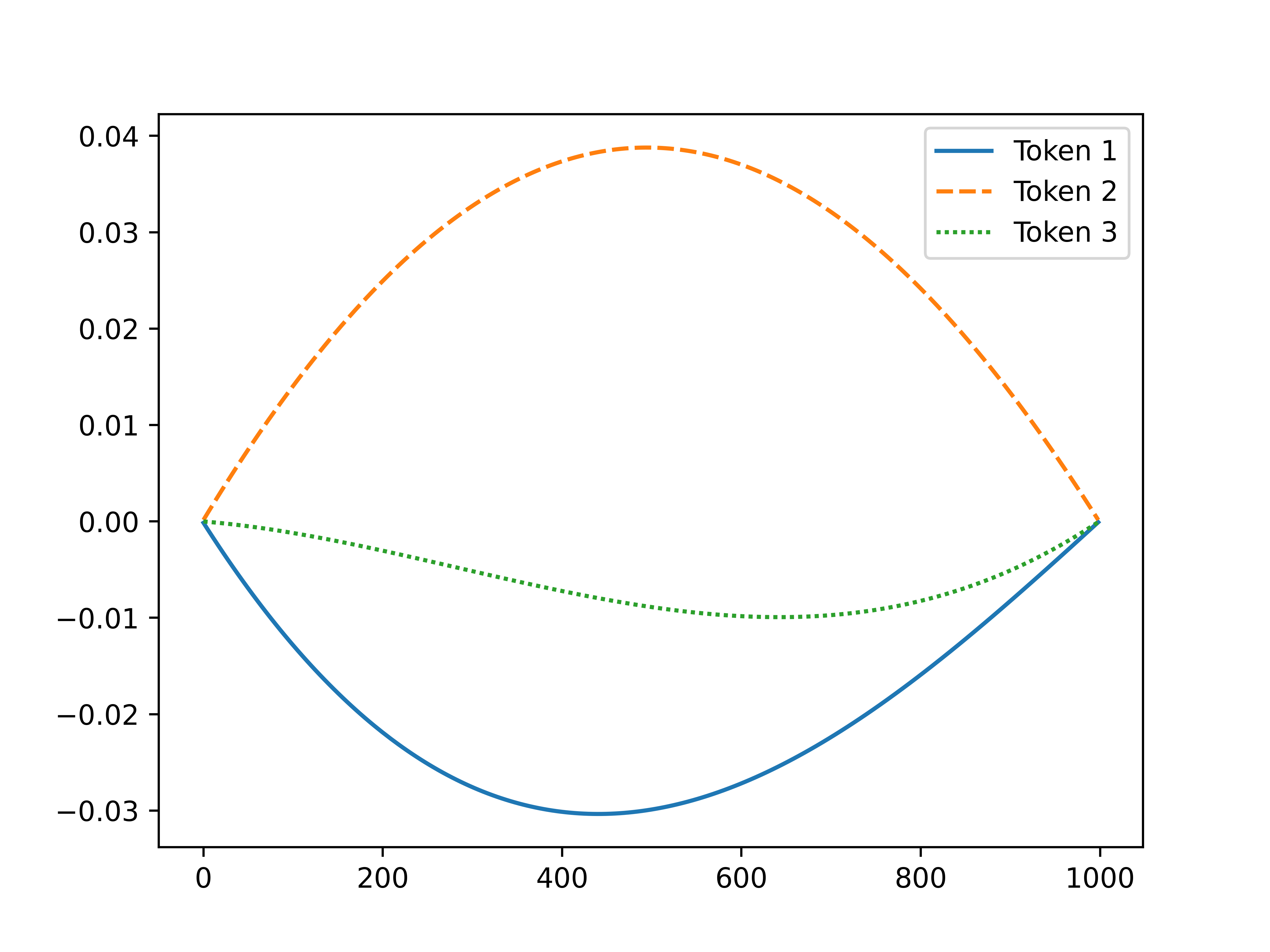}
    \caption{Optimal minus linear}
    \label{fig:linear_diff}
    \end{subfigure}
    \begin{subfigure}{0.45\columnwidth}
    \centering
    \includegraphics[trim={0 0mm 0 0mm},clip,width=0.93\textwidth]{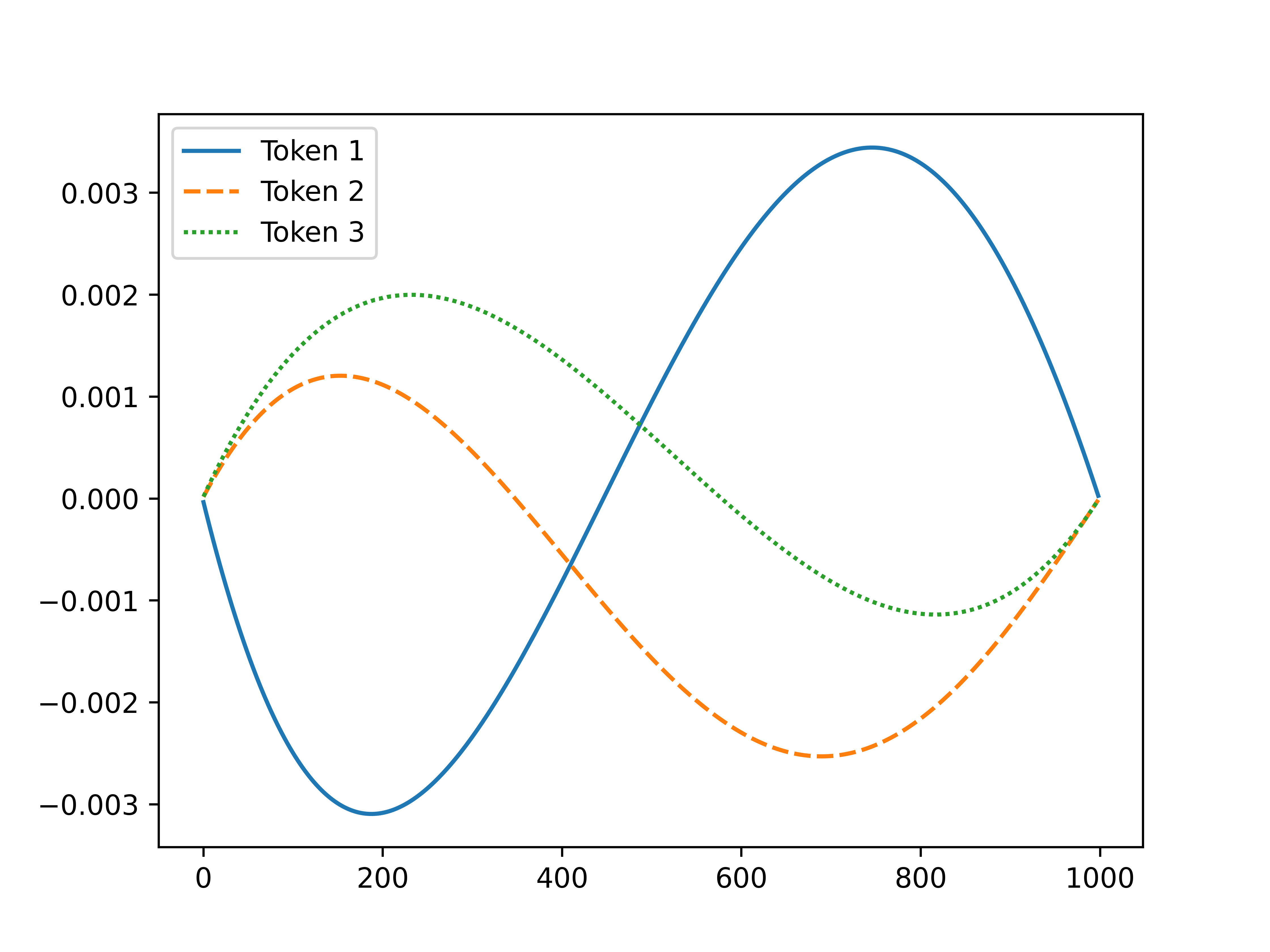}
    \caption{Optimal minus approximately-optimal}
    \label{fig:approx_diff}
    \end{subfigure}
    \label{fig:diff}
\end{figure}

\section{Experiments}

\subsection{Numerical Examples}
Here we find the optimal trajectory $\{\v w^*(t_k)\}_{k=1,...,f-1}$ by numerically solving Eq~\eqref{eq:opt_w_interp} for a given choice of $N$, $f$, $\v w(t_0)$ and $\v w(t_f)$.

In Figures~\ref{fig:traj} \&~\ref{fig:change}  we compare linear interpolation and $\{\breve{\v w}(t_k)\}_{k=1,...,f-1}$ and numerically-found weights $\{\v w^*(t_k)\}_{k=1,...,f-1}$ for a 3-token pool.
In Figure~\ref{fig:traj} we plot the weights themselves, in Figure~\ref{fig:change} we plot the block-to-block changes in weights.

Even though the overall weight change is large, $\v w(t_0)=\{0.05, 0.55, 0.4\}$ and $\v w(t_f)=\{0.4, 0.5, 0.1\}$, and we do this over $f=1000$ steps, our approximate method and the optimal are hard to visually distinguish.

In Figure~\ref{fig:diff} we plot the differences between numerically-found $\{\v w^*(t_k)\}_{k=1,...,f-1}$ and each of linear and approximately-optimal interpolations.
Clearly the approximately-optimal interpolation is much closer to optimal, with a largest absolute weight discrepancy of $\approx0.003$ compared to $\approx0.04$ for linear interpolation.

In this example the approximately-optimal trajectory $\{\breve{\v w}(t_k)\}_{k=1,...,f-1}$ captures $\approx95\%$ of the increase in pool value that one would get from swapping out linear interpolation with the numerical $\{\v w^*(t_k)\}_{k=1,...,f-1}$.
\newpage
\begin{figure}[h]
    \caption{Plots showing the trained returns for (a) approximately-optimal weight changes (b) linear weight changes. We plot the raw returns over the period. The performance increase from using approximately-optimal weight changes is maintained even as fees are introduced, even with fees of 1\%.}
    \centering
    \begin{subfigure}{0.45\columnwidth}
    \centering
    \includegraphics[trim={0 0mm 0 0mm},clip,width=0.93\textwidth]{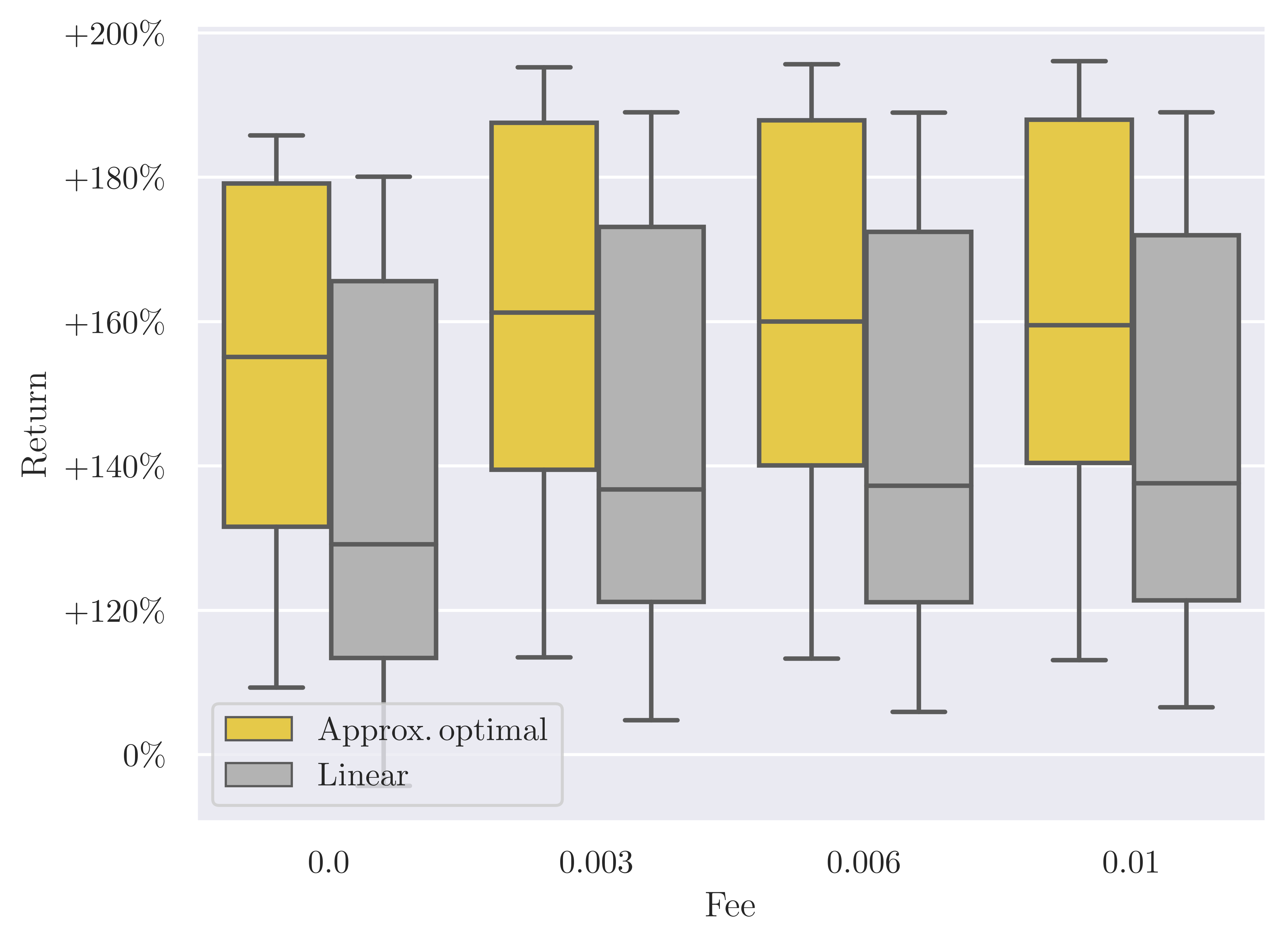}
    \caption{Momentum strategy}
    \label{fig:momentum}
    \end{subfigure}
    \begin{subfigure}{0.45\columnwidth}
    \centering
    \includegraphics[trim={0 0mm 0 0mm},clip,width=0.93\textwidth]{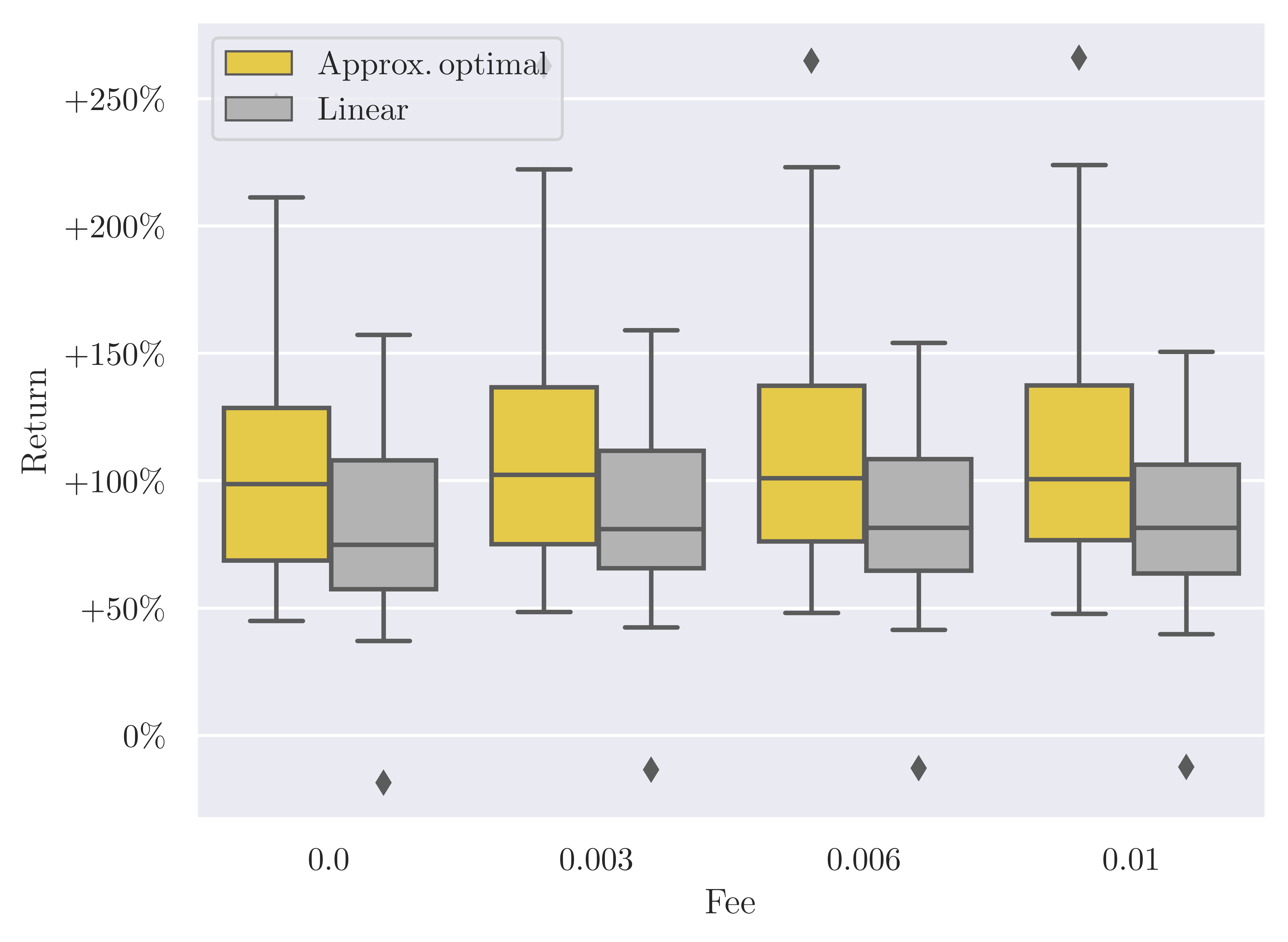}
    \caption{Channel following strategy}
    \label{fig:mean_rev}
    \end{subfigure}
    \label{fig:weight_change_performance}
\end{figure}
\subsection{Historical data}
We can train strategies for TFMMs, following the approach outlined the QuantAMM litepaper~\cite{quantamm_litepaper}.
New target weights are calculated using as input estimates of the gradients of the (log) prices of assets in the pool, for example, a simple momentum strategy.
These strategies have numerical parameters that control how the price gradients are calculated and how aggressively the pool acts to change weights based on those price gradients.
Different parameters lead to different performance.

Very briefly, we use stochastic gradient descent to tune a strategy.
First, we simulate the reserves held by a pool held over time from the interplay between the strategy's changing weights and changing market prices over a sampled period of price history.
From this we can calculate measures of the performance of the pool (e.g. Sharpe ratio) over that time period.
We perform this calculation in auto-differentiable computer code so we can take the gradient of the chosen performance measure with respect to the numerical parameters of the strategy.
We then update the parameters to increase measured performance.

The weight update procedure used (linear vs approximately-optimal) changes pool performance.
To most fairly test the effect of weight update procedure one has to train multiple different instances of strategies with each of these weight update procedures.
Instances of strategies learn to make use of the weight change procedure used.
A strategy trained under approximately-optimal weight changes might dramatically underperform if re-applied to rebalance via linear changes, but by separately training strategies that rebalance via linear changes we can then fairly compare performance.

We train 160 variants for each combination of strategy and weight change procedure, on a basket of BTC, ETH, DAI for the period July 2022 to June 2023.
The different variants calculate new target weights at different cadences (every hour vs. every day), start with different random initial parameters, and have slightly different methods of windowing data during training.

We train both a simple momentum strategy and a channel-following strategy.
We train with zero fees, but we run trained strategies over price data with fees present, following recent approaches for closed-form optimal arbitrage trades in multi-token pools~\cite{willetts2024closedform}.

We show the results, plotting raw returns in Figure~\ref{fig:weight_change_performance}.
We find that strategies can effectively learn to make use of approximately-optimal weight changes, giving increases in returns of $\sim25\%$ on average for the basket and period we have tested.

Further, this improvement is maintained (or even grows) as trading fees are introduced, showing the robustness of this approach.
Incidentally, one can think of comparing approximately-optimal weight changes with high trading fees to linear weight changes with low fees as a rough test of having a larger trading gas cost for approximately-optimal weight changes---as the trader might have to pay more for the weights to be loaded or for the calculation to be done on-chain during the trade.
(Of course on cheap L2s this effect would be minimal.)
See Figure~\ref{fig:weight_change_performance_alt} for the same data plotted to make this even easier to see.

\section{Concluding remarks}

Improved methods to dynamically change pool holdings are key to unlocking on-chain asset management for DeFi.
Doing this to reduce arbitrage opportunities during rebalancing leads to greater market efficiency.
We have introduced a new method for achieving an overall change in holdings that reduces rebalancing costs for a dynamic pool that rebalances its holdings by the construction of AMM-style arbitrage opportunities.
This approach can lead to increased returns in simulation and is robust to large trading fees being charged.

\newpage
\bibliography{biblio}

\newpage
\begin{appendices}
\renewcommand\thefigure{\thesection.\arabic{figure}}    
\renewcommand\theequation{\thesection.\arabic{equation}}

\section{Derivation of Optimal and Approximately-optimal weight change procedures}
\label{app:optimal}
\subsection{Optimal weight changes in the limit of small weight changes}
While we show below that a broad family of possible interpolations (\S\ref{app:general_interpolation}) and linear interpolations themselves (\S\ref{app:linear_interpolation}), can we do better?

Here we derive Eq~\eqref{eq:best_w_tilde}, the optimal intermediate weight in a 2-step process in the limit of small weight changes.

Taking partial derivatives of Eq~\eqref{eq:r_two_to_one_ratio}, we get
\begin{equation}
    \frac{\partial r}{\partial \Tilde{w}_i} = r \left(\frac{w_i(t_f)}{\Tilde{w}_i}+\log\left(\frac{w_i(t_0)}{\Tilde{w}_i}\right)-1\right).
\end{equation}
Solving for $\frac{\partial r}{\partial \Tilde{w}_i}=0$ gives us the trancendental equation
\begin{equation}
1-\log\left(\frac{w_i(t_0)}{\Tilde{w}_i}\right)=\frac{w_i(t_f)}{\Tilde{w}_i},
\label{eq:w_tilde_as_solution}
\end{equation}
which has solution
\begin{equation}
\Tilde{w}_i=\frac{w_i(t_f)}{W_0\left(\frac{e w_i(t_f)}{w_i(t_0)}\right)},
\label{eq:w_tilde_optimal}
\end{equation}
where $W_0(\cdot)$ is the principal branch of the function that solves the equation $w\exp{w}=x$, the Lambert W function.

As we have not required that $\sum_{i=1}^N \Tilde{w}_i=1$, Eq~\eqref{eq:w_tilde_optimal} is only valid in the limit $\lim_{|\v w(t_f)-\v w(t_0)|_1 \to 0}$.

\subsection{Bounding $\Tilde{w}_i$ above and below}
\label{app:bounding}
To avoid having to calculate $W_0(\cdot)$ on chain we instead bound $\Tilde{w}_i$ from above and below and then use the average of these bounds as an approximate value.

We will bound $\Tilde{w}_i$ from below by the geometric mean of the start and end weights, and from above by their arithmetic mean.

\paragraph{Bounding from below}
We will show
\[\Tilde{w}_i \geq \sqrt{w_i(t_0)w_i(t_f)}.\]

We have that $\Tilde{w}_i=\frac{w_i(t_f)}{W_0\left(\frac{e w_i(t_f)}{w_i(t_0)}\right)}$, which can be rearranged using the property $\frac{x}{W_0(x)}=\exp(W(x))$ to
\begin{equation}
    \Tilde{w}_i=w_i(t_0) \exp\left(W_0\left(\frac{e w_i(t_f)}{w_i(t_0)}\right)-1\right).
\end{equation}
Requiring $\Tilde{w}_i \geq \sqrt{w_i(t_0)w_i(t_f)}$ is equivalent to
\begin{align}
    &w_i(t_0) \exp\left(W_0\left(\frac{e w_i(t_f)}{w_i(t_0)}\right)-1\right)\geq\sqrt{w_i(t_0)w_i(t_f)}\\
    \Rightarrow&\exp\left(W_0\left(\frac{e w_i(t_f)}{w_i(t_0)}\right)-1\right)\geq\sqrt{\frac{w_i(t_f)}{w_i(t_0)}}\\
    \Rightarrow& W_0\left(\frac{e w_i(t_f)}{w_i(t_0)}\right)-1 \geq \frac{1}{2}\log\left(\frac{w_i(t_f)}{w_i(t_0)}\right).\label{eq:w_gm_inequality}
\end{align}
Defining $u:=\frac{w_i(t_f)}{w_i(t_0)}$ and $g(u):=W_0(e u)-1- \frac{1}{2}\log\left(u\right)$, we find that $g(u)$ has a single positive real root at $u=1$, which means that Eq~\eqref{eq:w_gm_inequality} reaches equality only for $\frac{w_i(t_f)}{w_i(t_0)}=1$.

It is easy to show that
\[\frac{\partial g}{\partial u}=\frac{1}{2}\frac{W_0(e u)-1}{u (1 + W_0(e u))}\]
We find that at the root we have
\[\Rightarrow \frac{\partial g}{\partial u}\bigg\rvert_{u=1}=0,\]
(where we have used the result $W_0\left(e\right)=1$).
Further,
\[\frac{\partial^2 g}{\partial u^2}=\frac{1}{2}\frac{1 + 3 W_0(e u) - \left(W_0(e u)\right)^2 - \left(W_0(e u)\right)^3}{u^2 (1 + W_0(e u))^3}\]
\[\Rightarrow \frac{\partial^2 g}{\partial u^2}\bigg\rvert_{u=1}=\frac{1}{8}.\]
As $g(u)$ has a single positive real root, which is a turning point with positive second derivative, $g(u)\geq0$ for all real values of $u>0$.
Thus $W_0\left(\frac{e w_i(t_f)}{w_i(t_0)}\right)-1 \geq \frac{1}{2}\log\left(\frac{w_i(t_f)}{w_i(t_0)}\right)$ and so $\Tilde{w}_i \geq \sqrt{w_i(t_0)w_i(t_f)}$ as required.

\paragraph{Bounding from above}
We will show that
\[\Tilde{w}_i \leq \frac{1}{2}\left({w_i(t_0)+w_i(t_f)}\right)\]
is satisfied under the requirement that weights have entries between 0 and 1.
Requiring $\Tilde{w}_i \leq \frac{1}{2}\left({w_i(t_0)+w_i(t_f)}\right)$ is equivalent to
\begin{align}
    &w_i(t_0) \exp\left(W_0\left(\frac{e w_i(t_f)}{w_i(t_0)}\right)-1\right)\leq \frac{1}{2}\left({w_i(t_0)+w_i(t_f)}\right)\\
    \Rightarrow&\exp\left(W_0\left(\frac{e w_i(t_f)}{w_i(t_0)}\right)-1\right)\leq \frac{1}{2}\left({1+\frac{w_i(t_f)}{w_i(t_0)}}\right)\\
    \Rightarrow& W_0\left(\frac{e w_i(t_f)}{w_i(t_0)}\right)-1 \leq \log\left(\frac{1}{2}\left({1+\frac{w_i(t_f)}{w_i(t_0)}}\right)\right).\label{eq:w_am_inequality}
\end{align}
Defining $u:=\frac{w_i(t_f)}{w_i(t_0)}$ and $h(u):=W_0(e u)-1- \log\left(\frac{1}{2}\left({1+u}\right)\right)$, we find that $h(u)$ has a single positive real root at $u=1$, which means that Eq~\eqref{eq:w_am_inequality} reaches equality only for $\frac{w_i(t_f)}{w_i(t_0)}=1$.

It is easy to show that
\[\frac{\partial h}{\partial u}=\frac{W_0(e u)}{u (1 + W_0(e u))}-\frac{1}{1+u}\]
We find that at the root we have
\[\Rightarrow \frac{\partial h}{\partial u}\bigg\rvert_{u=1}=0,\]
(where we have used the result $W_0\left(e\right)=1$).
Further,
\[\frac{\partial^2 h}{\partial u^2}=\frac{1}{u^2(1 + W_0(e u))}+\frac{1}{u^2(1 + W_0(e u))^2}-\frac{1}{u^2(1 + W_0(e u))^3}-\frac{2u+1}{u^2(1+u)^2}\]
\[\Rightarrow \frac{\partial^2 h}{\partial u^2}\bigg\rvert_{u=1}=-\frac{1}{8}.\]
As $h(u)$ has a single positive real root, which is a turning point with negative second derivative, $h(u)\leq0$ for all real values of $u>0$.
Thus $W_0\left(\frac{e w_i(t_f)}{w_i(t_0)}\right)-1 \leq \log\left(\frac{1}{2}\left({1+\frac{w_i(t_0)}{w_i(t_f)}}\right)\right)$ and so $\Tilde{w}_i \leq \frac{1}{2}\left({w_i(t_0)+w_i(t_f)}\right)$ as required.
\newpage
\subsubsection{Plots}
Here we plot the arithmetic \& geometric means of $w_i(t_0),w_i(t_f)$ and the optimal $\Tilde{w}_i$, showing the bounding we have derived.
\begin{figure}[h!]
    \centering
    \caption{Plots showing graphically how the bounds work and how they form an approximation to the optimal $\Tilde{w}_i$.}
    \begin{subfigure}{\columnwidth}
    \centering
    \includegraphics[width=7cm]{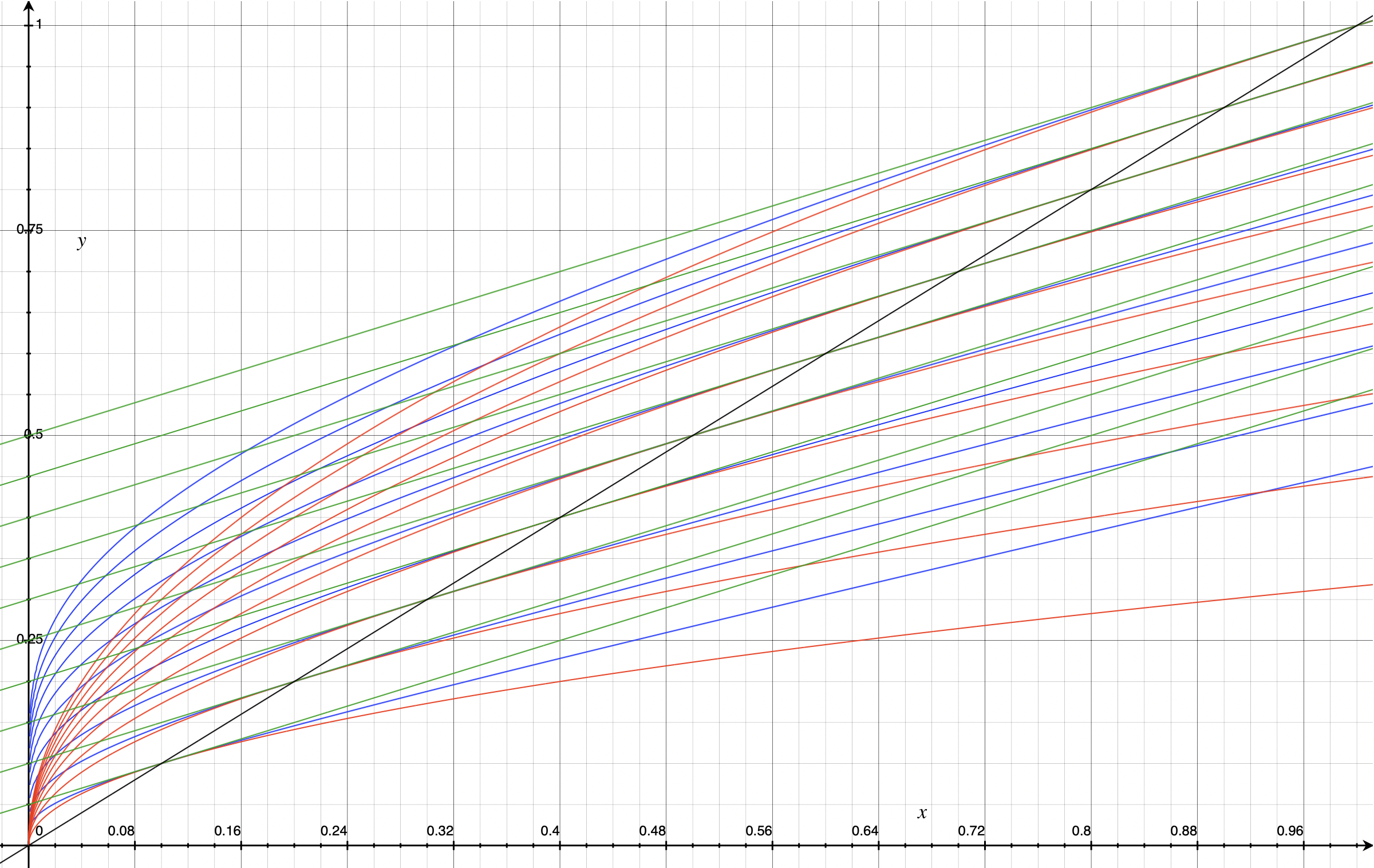}
    \caption{The arithmetic means (green) and geometric means (red) of $w_i(t_0),w_i(t_f)$, plotted as a function of $w_i(t_f)$ for 10 uniform values of $w_i(t_0)$ from 0.1 to 1.0. In blue are the optimal $\Tilde{w}_i$. Note that for each set of curves the optima (blue) is bounded from below by the geometric mean (red) and from above by the arithmetic mean (green). The black line is y=x; when $w_i(t_0)=w_i(t_f)$ the means and the optima all converge.}
    \label{fig:w_tilde_plots}
    \end{subfigure}
    \begin{subfigure}{\columnwidth}
    \centering
    \includegraphics[width=7cm]{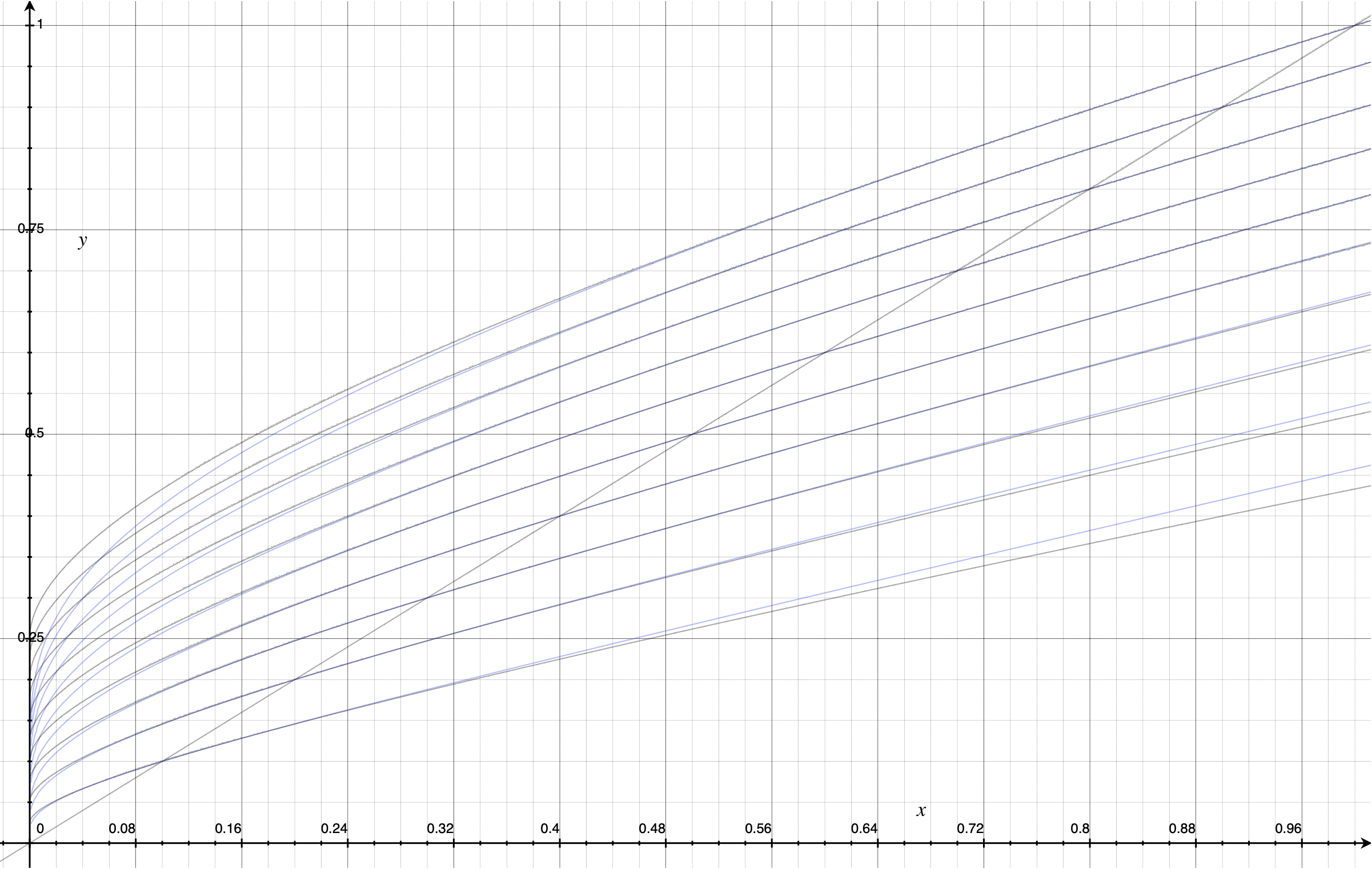}
    \caption{$\Tilde{w}_i$ (blue) plotted as a function of $w_i(t_f)$ for 10 uniform values of $w_i(t_0)$ from 0.1 to 1.0. The 10 black curves are the arithmetic means of the geometric and arithmetic means of $w_i(t_0),w_i(t_f)$ (the red and green curves above). This bound-average gives a close approximation to $\Tilde{w}_i$ and is most accurate near y=x (i.e. $w_i(t_0)\approx w_i(t_f)$).}
    \label{fig:w_tilde_plots_approx}
    \end{subfigure}
\end{figure}
\subsection{Approximately-optimal trajectories}
\label{app:bootstrap}
We bootstrap our `infinitesimal' results for optimal and approximately-optimal intermediate weights into a trajectory of many intermediates, Eqs~(\ref{eq:approx_optimal_traj_am}-\ref{eq:approx_optimal_traj}).
\begin{align}
    \breve{w}_i(t_k) &= \frac{w_i^{\mathrm{AM}}(t_k)+w_i^{\mathrm{GM}}(t_k)}{\sum_{j=1}^N\left({w_j^{\mathrm{AM}}(t_k)+w_j^{\mathrm{GM}}}\right)} \label{eq:app_approx_optimal_traj}\\
    w_i^{\mathrm{AM}}(t_k) & = (1-\frac{k}{f})w_i(t_0) + \frac{k}{f}w_i(t_f) \label{eq:app_approx_optimal_traj_am}\\
    w_i^{\mathrm{GM}}(t_k) & = {(w_i(t_0))^{(1-\frac{k}{f})}(w_i(t_f))^{\frac{k}{f}}},\label{eq:app_approx_optimal_traj_gm}
\end{align}
This works as locally-intermediate values of $w_i^{\mathrm{AM}}(t_k)$ and $w_i^{\mathrm{GM}}(t_k)$ are themselves the arithmetic and geometric means of their immediate neighbours:
\[w_i^{\mathrm{GM}}(t_k) = \sqrt{w_i^{\mathrm{GM}}(t_{k-1})w_i^{\mathrm{GM}}(t_{k+1}))}, \quad w_i^{\mathrm{AM}}(t_k) = \frac{1}{2}\left(w_i^{\mathrm{GM}}(t_{k-1+})w_i^{\mathrm{GM}}(t_{k+1}))\right).\]
So locally we have a sequence of relatively small weight changes, each $\breve{w}_i(t_k)$ of which closely approximates Eq~\eqref{eq:best_w_tilde}.
\section{Reserve-Update Derivation: Weights Change, Prices Constant}
\label{app:tfmm_update}
\emph{The result in this section previously appeared in the TFMM litepaper~\nbcite[Appendex A.2]{tfmm_litepaper}, we reproduce them here for clarity and ease of access.}

Here we derive Eq~\eqref{eq:reserve_change_weights}, the change in reserves of a pool where weights change but market prices are fixed.
For simplicity we are assuming the no-fees case.

At time $t_0$ we have weights $\v w(t_0)$ and a block later $t'=t_0 + \delta_t$ we have weights $\v w(t')$.

Just after the change in weights, we have the invariant of the pool
\begin{equation}
    \prod_{i=1}^N R_i(t_0)^{w_i(t')} = k(t'),
\end{equation}
and by the requirement that this quantity does not change under trading, we also have that after trading
\begin{equation}
    \prod_{i=1}^N R_i(t')^{w_i(t')} = k(t') =\prod_{i=1}^N R_i(t_0)^{w_i(t')}.
\end{equation}
\begin{equation}
   \Rightarrow \prod_{i=1}^N \left(\frac{R_i(t')}{R_i(t_0)}\right)^{w_i(t')}=1.
    \label{eq:k_tfmm_ratio}
\end{equation}

Meanwhile, we are assuming that arbitrageurs trade with the pool until the quoted prices match the market prices.
Prices are $\v p(t_0)$ before and after the change in weights.
Using standard results for quoted prices in G3Ms (for example \cite[Eq A.14]{quantamm_litepaper}), and without loss of generality using the first token as the num\'eraire, we have that initially, before the change in weights,
\begin{equation}
    \frac{\frac{w_i(t_0)}{R_i(t_0)}}{\frac{w_1(t_0)}{R_1(t_0)}} = p_i(t_0)
\end{equation}
and then after both updating the weights \& the pool reaching a new equilibrium
\begin{equation}
    \frac{\frac{w_i(t')}{R_i(t')}}{\frac{w_1(t')}{R_1(t')}} = p_i(t_0).
\end{equation}

Thus we have
\begin{align}
    \frac{w_i(t') w_1(t_0)}{R_i(t') R_1(t_0)}&=\frac{w_1(t') w_i(t_0)}{R_1(t') R_i(t_0)} \label{eq:w_tfmm},\\
    \frac{R_i(t')}{R_i(t_0)}&=\frac{w_i(t') w_1(t_0) R_1(t')}{w_1(t') w_i(t_0) R_1(t_0)} \label{eq:r_tfmm}.
\end{align}
Subbing Eq~\eqref{eq:r_tfmm} into Eq~\eqref{eq:k_tfmm_ratio}, we get
\begin{align}
    \prod_{i=1}^N \left(\frac{R_i(t')}{R_i(t_0)}\right)^{w_i(t')} &= \prod_{i=1}^N \left(\frac{w_i(t') w_1(t_0) R_1(t')}{w_1(t') w_i(t_0) R_1(t_0)}\right)^{w_i(t')} = 1 \\
    &=\frac{w_1(t_0) R_1(t')}{w_1(t') R_1(t_0)}\prod_{i=1}^N  \left(\frac{w_i(t')}{w_i(t_0)}\right)^{w_i(t')} =1\\
    &\Rightarrow \frac{w_1(t_0) R_1(t')}{w_1(t') R_1(t_0)} = \prod_{i=1}^N \left(\frac{w_i(t_0)}{w_i(t')}\right)^{w_i(t')} \label{eq:w1_ratio}.
\end{align}
Now subbing Eq~\eqref{eq:w1_ratio} into Eq~\eqref{eq:r_tfmm} and rearranging we get
\begin{equation}
    R_i(t') = R_i(t_0) \frac{w_i(t')}{w_i(t_0)}\prod_{j=1}^N \left(\frac{w_j(t_0)}{w_j(t')}\right)^{w_j(t')},
    \label{eq:apdx_weight_reserve_update}
\end{equation}
as we have in Eq~\eqref{eq:reserve_change_weights} in the main paper, completing this derivation.

\newpage

\section{Two-step updates are better than one-step updates}
\emph{Some of the result in this section previously appeared in the TFMM litepaper~\nbcite[Appendex A.3]{tfmm_litepaper}, we reproduce them here for clarity and ease of access.}

\subsection{A broad family of possible interpolations}
\label{app:general_interpolation}
We will compare two possible weight change methods, against a background of constant prices.
We have $\gamma=1$.

The first is the simple 1-step process, Eq~\eqref{eq:apdx_weight_reserve_update}, derived above.
Here we go directly from initial weights to final weights.
For compactness here, we will denote $\v w(t_0) + \Delta \v w$ as $\v w(t_f)$.
We have $\v w(t_0) \rightarrow \v w(t_f)$.

The second process is to go to an intermediate set of weights $\Tilde{\v{w}}$, be arbed to those weights, and then proceed to the final weights:
$\v w(t_0) \rightarrow \Tilde{\v{w}} \rightarrow \v w(t_f)$.
The restriction we place here on the elements of $\Tilde{\v{w}}$ is that $\forall i,\, \Tilde{w}_i \in [w_i(t_0),w_i(t_f)]$.
We enforce the standard requirements on weights, that $0<\Tilde{w}_i<1$ and $\sum_{i=1}^N \Tilde{w}_i=1$.

We will show that the two-step process is always superior as it leads to greater pool reserves.

\paragraph{One-step}
Eq~\eqref{eq:apdx_weight_reserve_update} directly gives us the ratio of the change of reserves,
\[\frac{R_i(t_f)}{R_i(t_0)} = r_i^{\mathrm{1-step}} =\frac{w_i(t_f)}{w_i(t_0)}\prod_{j=1}^N \left(\frac{w_j(t_0)}{w_j(t_f)}\right)^{w_j(t_f)}.\]
\paragraph{Two-step}
Applying Eq~\eqref{eq:apdx_weight_reserve_update} to the first leg, we get
\[r_i^{\mathrm{1st-step}} =\frac{\Tilde{w}_i}{w_i(t_0)}\prod_{j=1}^N \left(\frac{w_j(t_0)}{\Tilde{w}_j}\right)^{\Tilde{w}_j}.\]
And to the second leg
\[r_i^{\mathrm{2nd-step}} =\frac{w_i(t_f)}{\Tilde{w}_i}\prod_{j=1}^N \left(\frac{\Tilde{w}_j}{w_j(t_f)}\right)^{w_j(t_f)}.\]
This gives and overall change of reserves of
\[r_i^{\mathrm{2-step}} =r_i^{\mathrm{1st-step}} r_i^{\mathrm{2nd-step}}=\frac{w_i(t_f)}{w_i(t_0)}\prod_{j=1}^N \left(\frac{w_j(t_0)}{\Tilde{w}_j}\right)^{\Tilde{w}_j}\left(\frac{\Tilde{w}_j}{w_j(t_f)}\right)^{w_j(t_f)}.\]

\paragraph{Comparison}
The ratio $r_i^{\mathrm{2-step}} / r_i^{\mathrm{1-step}}$ is

\begin{align}
r_i^{\mathrm{2-step}} / r_i^{\mathrm{1-step}} = r &= \prod_{j=1}^N \left(\frac{w_j(t_0)}{\Tilde{w}_j}\right)^{\Tilde{w}_j}\left(\frac{\Tilde{w}_j}{w_j(t_f)}\right)^{w_j(t_f)}\left(\frac{w_j(t_f)}{w_j(t_0)}\right)^{w_j(t_f)}\\
& = \prod_{j=1}^N \left(\frac{w_j(t_0)}{\Tilde{w}_j}\right)^{\Tilde{w}_j}\left(\frac{\Tilde{w}_j}{w_j(t_0)}\right)^{w_j(t_f)}\\
\Rightarrow r&= \prod_{j=1}^N \frac{w_j(t_0)^{\Tilde{w}_j}}{w_j(t_0)^{w_j(t_f)}}\frac{\Tilde{w}_j^{w_j(t_f)}}{\Tilde{w}_j^{\Tilde{w}_j}}.
\label{eq:r_two_to_one_ratio}
\end{align}

If each term in this product is $>1$, then we will have shown that the two step process leads to greater final reserves than a one-step process.
Recall that $\forall i,\,w_i(t_0)<\Tilde{w}_i<w_i(t_f)$.
Let us now consider the $j^{\mathrm{th}}$ term of the product.
There are two cases to consider, either $w_i(t_f)>w_i(t_0)$ or $w_i(t_f)<w_i(t_0)$.

The $j^{\mathrm{th}}$ term in the product is 
\begin{equation}
    f_j=\frac{w_j(t_0)^{\Tilde{w}_j}}{w_j(t_0)^{w_j(t_f)}}\frac{\Tilde{w}_j^{w_j(t_f)}}{\Tilde{w}_j^{\Tilde{w}_j}}.
    \label{eq:jth_term}
\end{equation}

\paragraph{$w_j(t_f)>w_j(t_0)$:}

If $w_j(t_f)>w_j(t_0)$, then $w_j(t_f)>\Tilde{w}_j$ and $w_j(t_0)<\Tilde{w}_j$.
Thus $\Tilde{w}_j = a_j w_j(t_0)$ for some $a_j>1$.
Subbing into Eq~\eqref{eq:jth_term}, we get
\[f_j=\frac{w_j(t_0)^{\Tilde{w}_j}}{w_j(t_0)^{w_j(t_f)}}\frac{w_j(t_0)^{w_j(t_f)}}{w_j(t_0)^{\Tilde{w}_j}}\frac{a_j^{w_j(t_f)}}{a_j^{\Tilde{w}_j}}=a_j^{w_j(t_f)-\Tilde{w}_j}>1.\]

\paragraph{$w_j(t_f)<w_j(t_0)$:}

If $w_j(t_f)<w_j(t_0)$, then $w_j(t_f)<\Tilde{w}_j$ and $w_j(t_0)>\Tilde{w}_j$.
Thus $w_j(t_0) = b_j \Tilde{w}_j$ for some $b_j>1$.
Subbing into Eq~\eqref{eq:jth_term}, we get
\[f_j=\frac{\Tilde{w}_j^{\Tilde{w}_j}}{\Tilde{w}_j^{w_j(t_f)}}\frac{\Tilde{w}_j^{w_j(t_f)}}{\Tilde{w}_j^{\Tilde{w}_j}}\frac{b_j^{\Tilde{w}_j}}{b_j^{w_j(t_f)}}=b_j^{\Tilde{w}_j-w_j(t_f)}>1.\]

\paragraph{Summary}
Thus for either increasing elements, $w_j(t_f)>w_j(t_0)$, or decreasing elements, $w_j(t_f)<w_j(t_0)$, the terms in the product $r$ are always $>1$, therefore taking the two-step process is always superior, as required.

\subsection{Linear Interpolation}
\label{app:linear_interpolation}
Here we will derive Eq~\eqref{eq:arb-bisection}, finding the relationship between $\v R^{1{\text -}\mathrm{step}}$, the reserves in the pool when we have directly updated the weights from $\v w(t_0)\rightarrow\v w(t_0) +\Delta \v w$, a one-step process, and $\v R^{2{\text -}\mathrm{step}}$, the reserves when we have done this weight update via a two step process: $\v w(t_0) \rightarrow \v w(t_0) + \frac{1}{2}\Delta \v w \rightarrow \v w(t_0) + \Delta \v w$.

First Eq~\eqref{eq:apdx_weight_reserve_update} derived above directly gives us the one-step value:
\begin{align}
    R^{1{\text -}\mathrm{step}}_i &= R_i(t') \\
    &= R_i(t_0) \frac{w_i(t_0)+\Delta w_i}{w_i(t_0)}\prod_{j=1}^N \left(\frac{w_j(t_0)}{w_j(t_0)+\Delta w_j}\right)^{w_j(t_0)+\Delta w_j}.
\end{align}
Now let's consider the two-step process.
First we do half of the update,
\begin{equation}
    R^{\Delta w/2}_i = R_i(t_0) \frac{w_i(t_0)+\frac{\Delta w_i}{2}}{w_i(t_0)}\prod_{j=1}^N \left(\frac{w_j(t_0)}{w_j(t_0)+\frac{\Delta w_j}{2}}\right)^{w_j(t_0)+\frac{\Delta w_j}{2}}.
\end{equation}
And then completing this two-step process
\begin{align}
    R^{2{\text -}\mathrm{step}}_i &= R^{\Delta w/2}_i \frac{w_i(t_0)+{\Delta w_i}}{w_i(t_0)+\frac{\Delta w_i}{2}}\prod_{j=1}^N \left(\frac{w_j(t_0)+\frac{\Delta w_j}{2}}{w_j(t_0)+{\Delta w_j}}\right)^{w_j(t_0)+\Delta w_j}\\
    &= R_i(t_0) \frac{w_i(t_0)+{\Delta w_i}}{w_i(t_0)}\prod_{j=1}^N \frac{w_j(t_0)^{w_j(t_0)+\frac{\Delta w_j}{2}}\left(w_j(t_0)+\frac{\Delta w_j}{2}\right)^\frac{\Delta w_j}{2}}{\left(w_j(t_0)+{\Delta w_j}\right)^{w_j(t_0)+{\Delta w_j}}}.
\end{align}
Dividing $R^{2{\text -}\mathrm{step}}_i$ by $R^{1{\text -}\mathrm{step}}_i$ we get
\begin{align}
    \frac{R^{2{\text -}\mathrm{step}}_i}{R^{1{\text -}\mathrm{step}}_i} &= \prod_{j=1}^N \left(1+\frac{\Delta w_j}{2w_j(t_0)}\right)^{\frac{\Delta w_j}{2}}.
\end{align}
Thus we have that
\begin{align}
    \v R^{2{\text -}\mathrm{step}} = \v R^{1{\text -}\mathrm{step}}\prod_{j=1}^N \left(1+\frac{\Delta w_j}{2 w_j(t_0)}\right)^{\frac{\Delta w_j}{2}},
\end{align}
as required.

\section{Supplementary plots}
\label{app:supp_plots}
\begin{figure}[h]
    \caption{Differing from Figure~\ref{fig:weight_change_performance}, we group results here by weight change method to make it easier to see the effect of trading fees. We show the trained returns for (a) approximately-optimal weight changes (b) linear weight changes. We plot the raw returns over the period. The performance increase from using approximately-optimal weight changes is maintained even as fees are introduced, even with fees of 1\%.}
    \centering
    \begin{subfigure}{0.45\columnwidth}
    \centering
    \includegraphics[trim={0 0mm 0 0mm},clip,width=0.93\textwidth]{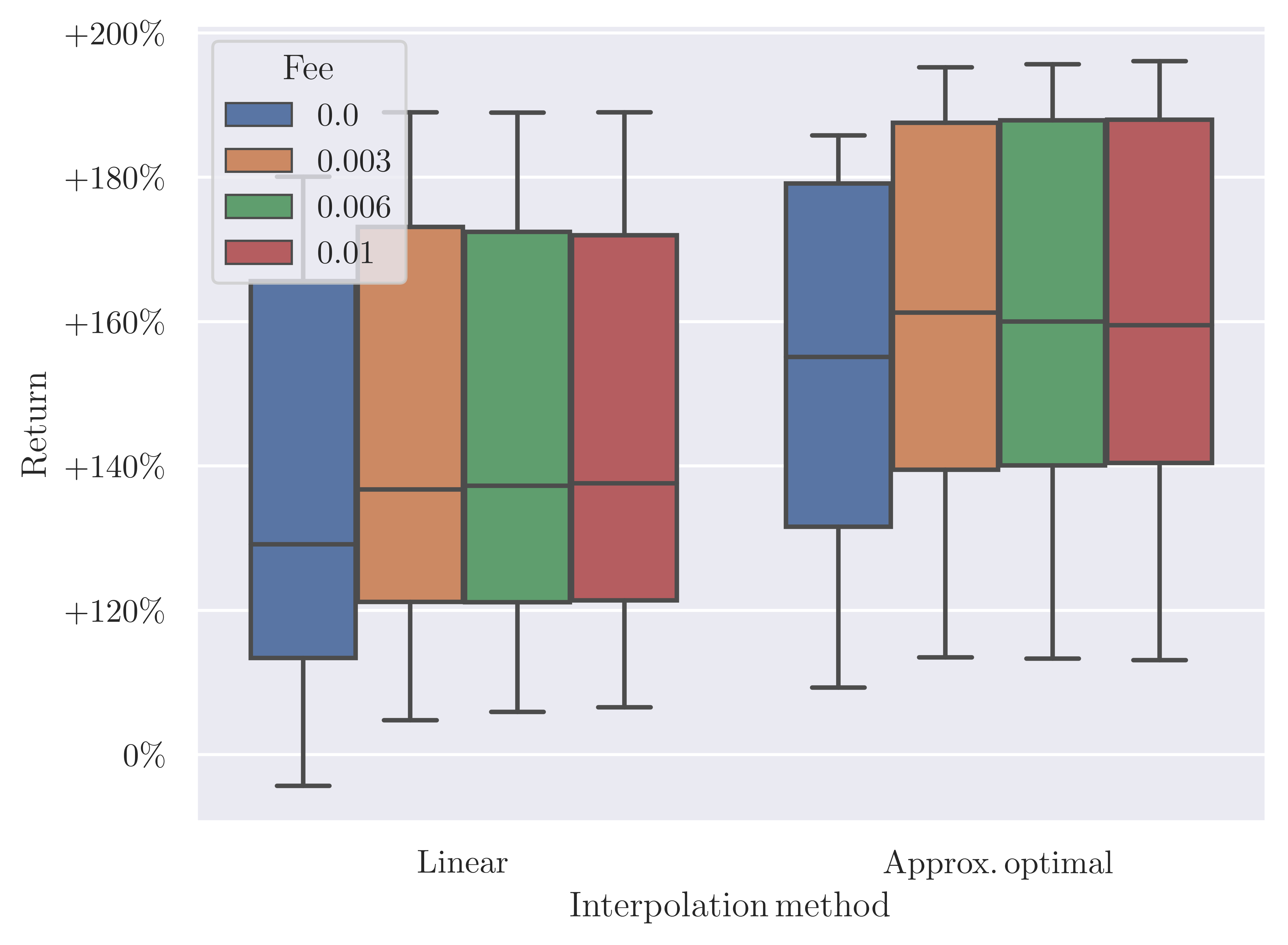}
    \caption{Momentum strategy}
    \label{fig:momentum_alt}
    \end{subfigure}
    \begin{subfigure}{0.45\columnwidth}
    \centering
    \includegraphics[trim={0 0mm 0 0mm},clip,width=0.93\textwidth]{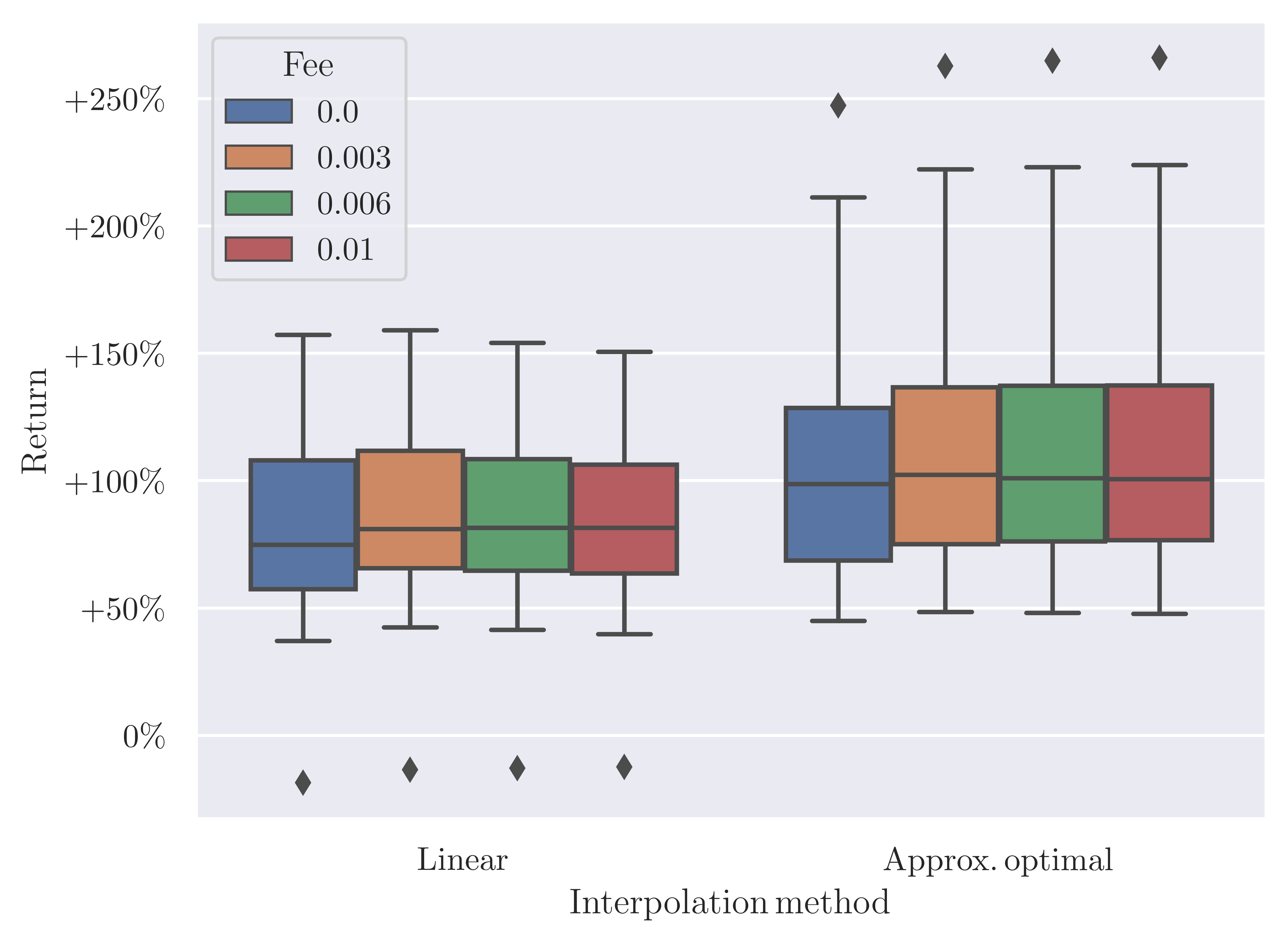}
    \caption{Channel following strategy}
    \label{fig:mean_rev_alt}
    \end{subfigure}
    \label{fig:weight_change_performance_alt}
\end{figure}

\end{appendices}

\newpage
\textbf{DISCLAIMER} This paper is for general information purposes only.
It does not constitute investment advice or a recommendation or solicitation to buy or sell any investment or asset, or participate in systems that use TFMM.
This paper should not be used in the evaluation of the merits of making any investment decision.
It should not be relied upon for accounting, legal or tax advice or investment recommendations.
This paper reflects current opinions of the authors regarding the development and functionality of TFMM and is subject to change without notice or update.

While some aspects, such as altering target weights in geometric mean market makers is prior art, aspects of TFMM that are novel such as, but not exclusively, composability mechanisms, efficient methods for gradients and covariances, generic form multi-token amplification and advanced execution management mechanisms for use in dynamic weight AMMs, for purposes of core liquidity providing or forms of asset management including, but not exclusively, fund construction, structured products, treasury management are covered by patent filing date of 21st February 2023.

\end{document}